\documentclass[10pt]{iopart}
\usepackage{latexsym,hyperref,color,amssymb,graphicx}
\newcommand{\half}{\frac{1}{2}}
\newcommand{\thalf}{{\textstyle{\frac{1}{2}}}}
\newcommand{\shalf}{{\scriptstyle{ \frac{1}{2}}}}

\newcommand{\tcar}{{\textstyle{\frac{1}{4}}}}

\newcommand{\beq}{\begin{equation}}
\newcommand{\eeq}{\end{equation}}
\newcommand{\bea}{\begin{eqnarray}}
\newcommand{\eea}{\end{eqnarray}}
\newcommand{\bean}{\begin{eqnarray*}}
\newcommand{\eean}{\end{eqnarray*}}
\newcommand{\bei}{\begin{itemize}}
\newcommand{\eei}{\end{itemize}}
\newcommand{\ben}{\begin{enumeration}}
\newcommand{\een}{\end{enumeration}}
\newcommand{\nn}{\nonumber}

\definecolor{darkorange}{rgb}{.6,.2,.0}

\definecolor{darkgreen}{rgb}{0.0,0.7,0.0}

\begin{document}
\title{Learning about Spin-One-Half Fields}
\author{Peter Cahill}
\address{Department of Mechanical Engineering, 
University of New Mexico, Albuquerque, NM 87131}
\ead{pedo@unm.edu}
\author{Kevin Cahill}
\address{New Mexico Center for Particle Physics, 
Department of Physics and Astronomy, 
University of New Mexico, Albuquerque, NM 87131}
\ead{cahill@unm.edu}
\date{\today}
\begin{abstract}
It is hard to understand
spin-one-half fields without
reading Weinberg.
This paper is a pedagogical
footnote to his formalism
with an emphasis on the boost matrix,
spinors, and Majorana fields.
\end{abstract}
\maketitle
\section{Introduction}
The construction of Majorana and Dirac fields 
confuses many students year after year.
Those who are not confused are likely
to be those who have learned the subject
from Weinberg's 
papers~\cite{WeinbergFRAS,WeinbergFRAS:M=0,WeinbergFRAS:III} 
or from volume one~\cite{WeinbergI} 
of his treatise
on quantum field theory.
In that book,
he describes how states respond to
Poincar\'{e} transformations and 
infers how creation and annihilation
operators transform.
Then he shows that fields,
which are linear combinations
of these operators, 
transform suitably if their coefficients,
the spinors, are related to suitable
zero-momentum spinors by a ``standard boost''
matrix \( D(L(p)) \)\@.
We cannot improve upon Weinberg's treatment,
but we can add to it ---
the present paper is a long
pedagogical footnote to section 5.5
of his book~\cite{WeinbergI}\@.
\par
In Section~\ref{RCMFS}, we recall his 
construction of spinors from
the Dirac representation \( D(L(p)) \)
of the standard boost \( L(p) \)\@.
In Section~\ref{S}, we derive 
for the matrix \( D(L(p)) \) the 
simple expression 
\beq
D(L(p)) = 
\frac{
\left( m + p_a \gamma^a \, 
\gamma^0 \right)}
{\sqrt{2m(p^0+m)}}
\label {introBoost}
\eeq
which leads directly to 
useful matrix formulas for 
the spinors \(u(\mathbf{p},s)\)
and  \(v(\mathbf{p},s)\)\@.
We suggest classroom use of this expression
and these matrix formulas.
They imply
that the spinors satisfy the Dirac equation
in momentum space and the Majorana conditions,
and they simplify the evaluation of spin sums.
\par
In Section~\ref{MF},
we construct the Majorana field \( \chi(x) \)
and show that it satisfies the Majorana condition.
We use spin sums to compute its propagator
and its anti-commutator with itself and its adjoint.
We also relate it to a four-component, anti-commuting
scalar-like field.
In Section~\ref{DF}, we construct the Dirac field
\( \psi(x) \) from two Majorana fields 
that describe particles of the same mass,
and we show that its anti-particle field is
\( \psi_c = \gamma^2 \psi^* \) because
its constituent Majorana fields satisfy the
Majorana condition.
We discuss the propagator and the 
causality and helicity properties
of the Dirac field, which we relate 
to a complex, four-component, anti-commuting
scalar-like field.
Finally, in Section~\ref{susy},
we apply our Majorana formulas to the
Wess-Zumino model and its supercharges.
The lessons of sections (\ref{MF}--\ref{susy}) 
are appropriate for classroom use.
\section{Relativity, Causality, Majorana Fields, 
and Spinors\label{RCMFS}}
In this section, we present a distillation 
for Majorana fields
of Weinberg's discussion~\cite{WeinbergI} 
of spin-one-half fields.  This section provides
the context of this paper.
\par
The particle-annihilation 
\beq
\chi^+_b(x) = \int \!\! \frac{d^3p}{(2\pi)^{3/2}}
\sum_s
u_b(\mathbf{p},s) \, a(\mathbf{p},s) e^{ipx}
\label {Maj+}
\eeq
and particle-creation fields
\beq
\chi_b^-(x) = \int \!\! \frac{d^3p}{(2\pi)^{3/2}}
\sum_s
v_b(\mathbf{p},s) \, a^\dagger(\mathbf{p},s) e^{-ipx}
\label{Maj-} 
\eeq
of a spin-one-half Majorana \(\chi(x)\) 
field will transform suitably
under Poincar\'{e} transformations
\beq
U(\Lambda,a) \chi_a^\pm(x) U^{-1}(\Lambda,a) =
\sum_b D_{ab}(\Lambda^{-1}) \chi_b^\pm(\Lambda x + a)
\label {poincare}
\eeq
if the spinors \(u(\mathbf{p},s)\) 
and \(v(\mathbf{p},s)\) are related 
to suitable zero-momentum spinors 
\( u(\mathbf{0},s)\) and 
\(v(\mathbf{0},s) \) 
by a matrix \(D_{ab}(L(p))\)
\bea
u(\mathbf{p},s) & = & 
\sqrt{m/p^0} \, D(L(p)) \, u(\mathbf{0},s) \nn\\
v(\mathbf{p},s) & = & 
\sqrt{m/p^0} \, D(L(p)) \, v(\mathbf{0},s) 
\label {uvboost}
\eea
that represents the standard boost 
\beq
L(p)_b^{\;\;0} = - p_b/m
\label {sboost}
\eeq
that takes \((m,\mathbf{0})\) into \((p^0,\mathbf{p})\)
in the Dirac representation of the Lorentz group,
as explained in section 5.4 of ref.~\cite{WeinbergI}\@.
The \(4 \times 4\) matrices \( D(\Lambda)  \)
of this representation
are exponentials
\beq
D(\Lambda) = 
e^{\frac{i}{2} \omega_{ab} \mathcal{J}^{ab}}
\label {D0}
\eeq
of generators of the Lorentz group
\beq
\mathcal{J}^{ab} = - \frac{i}{4} [ \gamma^a, \gamma^b ]
\label {D}
\eeq
in which the gamma matrices 
satisfy the anti-commutation relations
\beq
\{ \gamma^a, \gamma^b \} = 2 \eta^{ab}
\label {AC}
\eeq
where \( \eta \) is the flat space-time metric 
\( \eta = \mbox{diag}(-1,1,1,1) \)\@.
In the Dirac representation, 
the gamma matrices \( \gamma^a \) 
transform as a vector
\beq
D(\Lambda) \gamma^a D^{-1}(\Lambda) =
\Lambda_b^{\; a} \gamma^b .
\label {gammaVec}
\eeq
\par
The fields \( \chi^\pm(x) \) will transform 
suitably under parity
if the  zero-momentum spinors 
\( u(\mathbf{0},s)\) and 
\(v(\mathbf{0},s) \) are eigenstates of \(i\gamma^0\)
with eigenvalues \(\pm 1\)\@.
\par
Although the fields \( \chi^\pm_b(x) \) do not
commute or anti-commute with their adjoints,
the Majorana field that is their sum
\beq
\!\!\!\!\!\!\!\!\!\!\!\!\!\!
\chi_b(x) = \int \!\! \frac{d^3p}{(2\pi)^{3/2}}
\sum_s
\left[ 
u_b(\mathbf{p},s) \, a(\mathbf{p},s) e^{ipx} 
+     
v_b(\mathbf{p},s) \, a^\dagger(\mathbf{p},s) e^{-ipx}
\right]
\label {MajField0} 
\eeq
will anti-commute at space-like
separations both with itself and its adjoint if
the operators \(a(\mathbf{p},\pm {\thalf})\) and
\(a^\dagger (\mathbf{p},\pm {\thalf})\) 
obey
the anti-commutation relations
\beq
\{ a(\mathbf{p}, s), a^\dagger (\mathbf{p'}, s') \}
= \delta_{s s'} \, \delta( \mathbf{p} - \mathbf{p'} )
\label {[aa*]}
\eeq
and
\beq
\{ a(\mathbf{p}, s), a(\mathbf{p'}, s') \} = 0
\label {[aa]}
\eeq
and if
the zero-momentum spinors are eigenstates
of \(i \gamma^0\) with opposite eigenvalues
(1 \& \(-1\) or \(-1\) \& 1)~\cite{WeinbergI}\@.
The usual choice is
\bea
i \gamma^0 u(\mathbf{0},s) & = & 
u(\mathbf{0},s) \nn\\
i \gamma^0 v(\mathbf{0},s) & = & \mbox{}
- v(\mathbf{0},s).
\label {igamma0}
\eea
If the gamma matrices
are taken to be
\beq
\gamma^k = -i \pmatrix{0&\sigma_k\cr
                       -\sigma_k&0}
\quad k = 1, 2, 3
\label {gamma123}
\eeq
and
\beq 
\gamma^0 = -i \beta 
= -i \pmatrix{\mathbf{0}&I\cr
                       I&\mathbf{0}},
\label {gamma0}
\eeq
then a natural choice~\cite{WeinbergI} 
for the zero-momentum spinors is
\beq
u(\mathbf{0},{\thalf}) = \frac{1}{\sqrt{2}}
\pmatrix{1\cr 
         0\cr
         1\cr 
         0\cr}, \qquad
u(\mathbf{0},-{\thalf}) = \frac{1}{\sqrt{2}}
\pmatrix{0\cr 
         1\cr
         0\cr 
         1\cr},
\label{u(0)}
\eeq
and
\beq
v(\mathbf{0},{\thalf}) = \frac{1}{\sqrt{2}}
\pmatrix{0\cr 
         1\cr
         0\cr 
        -1\cr}, \qquad
v(\mathbf{0},{-\thalf}) = \frac{1}{\sqrt{2}}
\pmatrix{-1\cr 
          0\cr
          1\cr 
          0\cr}
\label{v(0)}
\eeq
which incidentally 
satisfy the Majorana conditions
\bea
u(\mathbf{0},s) & = & \gamma^2 v^*(\mathbf{0},s) 
=  \gamma^2 v(\mathbf{0},s) \nn\\
v(\mathbf{0},s) & = & \gamma^2 u^*(\mathbf{0},s)
= \gamma^2 u(\mathbf{0},s) .
\label {MajCond0}
\eea
The spatial gamma matrices are hermitian,
and the temporal one is anti-hermitian
\beq
(\gamma^k)^\dagger = \gamma^k \quad
\mbox{and} \quad (\gamma^0)^\dagger = - \gamma^0.
\label {HC}
\eeq
The gamma matrices of even index are symmetric,
and those of odd index are anti-symmetric
\beq
\left( \gamma^a \right)^\top = (-1)^a \, \gamma^a 
= - \gamma^0 \gamma^2 \gamma^a  \gamma^0 \gamma^2 .
\label {transpose}
\eeq
The matrix \(\gamma_5\) is
\beq
\gamma_5 = -i \gamma^0 \gamma^1 \gamma^2 \gamma^3
= \pmatrix{I&0\cr
           0&-I}.
\label {gamma5}
\eeq
\par
With a different set of \(\gamma\)-matrices
\(
\gamma^{a '} = S \gamma^a S^{-1}, 
\)
the fields and spinors
should be multiplied from the left
by the matrix \(S\)\@.
A nice feature of the chosen \(\gamma\)-matrices
(\ref{gamma123} \& \ref{gamma0}) is that the
Lorentz generators \(\mathcal{J}^{ab}\)
are block diagonal, as is formula (\ref{introBoost})
for the standard boost 
\beq
D(L(p)) = \frac{1}{\sqrt{2m(p^0+m)}} \,
\pmatrix{p^0 + m - \mathbf{p} \cdot \vec \sigma&0\cr
         0&p^0 + m + \mathbf{p} \cdot \vec \sigma}
\label {bdSBOOST}
\eeq
in the Dirac representation of the Lorentz group.
A derivation of formula (\ref{introBoost})
for the standard boost is given in the next section.
\section{Spinors \label {S}}
In this section, we'll find useful matrix formulas for
the spinors \( u(\mathbf{p},s) \) and
\( v(\mathbf{p},s) \) from their
definitions (\ref{uvboost})\@.
The key step will be the explicit evaluation
of the standard boost \(D(L(p))\)\@.
We then use these matrix formulas 
to evaluate spin sums and to show
that the spinors obey Majorana conditions
and the Dirac equation in momentum space. 
\par
The standard boost \(D(L(p))\) is~\cite{WeinbergI}
\beq
D(L(p)) = 
D(R(\hat\mathbf{p}) \, D(B(|\mathbf{p}|)) \, 
D(R^{-1}(\hat\mathbf{p}) 
\label {D(L(p))}
\eeq
where \( B(|\mathbf{p}|) \) is a boost in the 
three-direction, and \( R(\hat\mathbf{p}) \) is a 
particular rotation that takes the three-axis
into the direction \(\hat\mathbf{p}\)\@.
Thus the standard boost is a boost in the
direction \(\hat\mathbf{p}\) that takes
the four-vector \((m,\mathbf{0})\) to
\(p\)\@.  The generator of such boosts is 
proportional to 
\(4i \mathcal{J}^{i0} \, p^i = 
[\gamma^i,\gamma^0] \, p^i\) and so to 
\(\hat\mathbf{p} \cdot \vec\mathbf{\gamma} \, \gamma^0\)\@.
So the standard boost \(D(L(p))\) is
\beq
D(L(p)) = 
e^{\alpha \hat\mathbf{p} \cdot \vec\mathbf{\gamma} 
\, \gamma^0}
\label {boost}
\eeq
in which \(\alpha \) is a parameter whose value
is constrained by the requirements 
(\ref{sboost} \& \ref{gammaVec})
\bea
D(L(p)) \gamma^0 D^{-1} (L(p)) & = & 
e^{\alpha \hat\mathbf{p} \cdot 
\vec\mathbf{\gamma} \, \gamma^0}
\gamma^0
e^{- \alpha \hat\mathbf{p} 
\cdot \vec\mathbf{\gamma} \, \gamma^0}
= L(p)_b^{\;\;0} \gamma^b \nn\\
& = & - p_b \gamma^b/m = - p_b \gamma^b /m .
\label {keyboost}
\eea
Because 
\( (\alpha \hat\mathbf{p} \cdot 
\vec\mathbf{\gamma} \, \gamma^0 )^2
= \alpha^2 \),
the gamma matrix 
\(\gamma^0\) is transformed to
\bea
e^{\alpha \hat\mathbf{p} \cdot 
\vec\mathbf{\gamma} \, \gamma^0}
\gamma^0
e^{- \alpha \hat\mathbf{p} \cdot 
\vec\mathbf{\gamma} \, \gamma^0}
& = &e^{2 \alpha \hat\mathbf{p} \cdot 
\vec\mathbf{\gamma} \, \gamma^0}
\gamma^0 \nn\\
& = & \left( \cosh 2\alpha + 
\hat\mathbf{p} \cdot \vec\mathbf{\gamma} \gamma^0
\sinh 2\alpha \right) \gamma^0 \nn\\
& = & \gamma^0 \cosh 2\alpha - 
\hat\mathbf{p} \cdot \vec\mathbf{\gamma} \sinh 2\alpha 
= - p_b \gamma^b /m 
\label {ch&sh}
\eea
by the preceding equation.
So \(\cosh 2 \alpha = p^0/m \), whence
\beq
\cosh \alpha = \sqrt{\frac{p^0+m}{2m}}
\label {chalpha}
\eeq
and
\beq
\sinh \alpha = \sqrt{\frac{p^0-m}{2m}}.
\label {shalpha}
\eeq
\par
Thus the standard boost \( D(L(p)) \) is
\bea
D(L(p)) & = &
e^{\alpha \hat\mathbf{p} \cdot 
\vec\mathbf{\gamma} \, \gamma^0}
= \cosh \alpha + 
\hat\mathbf{p} \cdot 
\vec\mathbf{\gamma} \gamma^0 \sinh \alpha
\nn\\
& = & \sqrt{\frac{p^0+m}{2m}} \, + \,
\hat\mathbf{p} \cdot \vec\mathbf{\gamma} \, \gamma^0 \,
\sqrt{\frac{p^0-m}{2m}} \nn\\
& = & \frac{
\left( p^0 + m + \mathbf{p} \cdot \vec \gamma \, 
\gamma^0 \right)}
{\sqrt{2m(p^0+m)}}
\label {SBOOST0}
\eea
or since \( \left(\gamma^0\right)^2 = - 1 \)
%which we may write as
\beq
D(L(p)) = 
\frac{
\left( m + p_a \gamma^a \, 
\gamma^0 \right)}
{\sqrt{2m(p^0+m)}} .
\label {SBOOST}
\eeq
%since \( \left(\gamma^0\right)^2 = - 1 \)\@.
This simple, explicit formula for the 
standard boost leads directly to expressions for
the spinors and their spin sums, and so
we recommend its use in classrooms.
It is true for any choice of gamma matrices
that satisfy \( \{ \gamma^a , \gamma^b \} = 2 \eta^{ab} \)
with \( \eta = \mbox{diag}(-1,1,1,1) \)\@.
\par
By its definition (\ref{uvboost}) in terms
of the standard boost (\ref{SBOOST}),
the spinor \( u(\mathbf{p},s) \) is
\beq
u(\mathbf{p},s) = 
\sqrt{\frac{m}{p^0}} \, D(L(p)) \, u(\mathbf{0},s) 
= \frac{
\left( m + p_a \gamma^a \, 
\gamma^0 \right)}
{\sqrt{2p^0(p^0+m)}} \, u(\mathbf{0},s)
\label {u(p,s)1}
\eeq
or
\beq
u(\mathbf{p},s) = 
 \frac{
\left( m - ip_a \gamma^a \right)}
{\sqrt{2p^0(p^0+m)}} \, u(\mathbf{0},s)
\label {u(p,s)}
\eeq
since 
\( \gamma^0 u(\mathbf{0},s) = -i u(\mathbf{0},s) \) 
by (\ref{igamma0})\@.
\par
Similarly, by its definition (\ref{uvboost})
in terms of the standard boost (\ref{SBOOST}),
the spinor \( v(\mathbf{p},s) \) is
\beq
v(\mathbf{p},s) = 
\sqrt{\frac{m}{p^0}} \, D(L(p)) \, v(\mathbf{0},s) 
= \frac{
\left( m + p_a \gamma^a \, 
\gamma^0 \right)}
{\sqrt{2p^0(p^0+m)}} \, v(\mathbf{0},s)
\label {v(p,s)1}
\eeq
or
\beq
v(\mathbf{p},s) = 
 \frac{
\left( m + ip_a \gamma^a \right)}
{\sqrt{2p^0(p^0+m)}} \, v(\mathbf{0},s)
\label {v(p,s)}
\eeq
since 
\( \gamma^0 v(\mathbf{0},s) = i v(\mathbf{0},s) \) 
by (\ref{igamma0})\@.
\par
These matrix formulas 
(\ref{u(p,s)} \& \ref{v(p,s)}) for
the spinors  \( u(\mathbf{p},s) \) and
\( v(\mathbf{p},s) \) follow from
the definition (\ref{uvboost}) 
of the spinors in terms of the standard boost,
the expression (\ref{SBOOST}) for that boost,
and from the eigenvalue equations (\ref{igamma0})\@.
They are quite
independent of the choice of gamma matrices.
For a different set of \(\gamma\)-matrices
\(
\gamma'^{\, a} = S \gamma^a S^{-1}, 
\)
one merely multiplies the 
\(\mathbf{p=0}\) spinors (\ref{u(0)} \& \ref{v(0)})
and the spinors \(u(\mathbf{p},s)\) and
\(v(\mathbf{p},s)\) of Eqs.(\ref{u(p,s)} \& \ref{v(p,s)})
by the matrix \(S\), an operation under which
the conditions 
(\ref{igamma0} \& \ref{MajCond0}) are covariant.
Our derivation of the boost formula
(\ref{SBOOST}) and of the matrix formulas 
(\ref{u(p,s)} \& \ref{v(p,s)}) and of some of 
their consequences are the main content
of this paper.
\par
The adjoints \(\overline{u}(\mathbf{p},s) 
= i u^\dagger(\mathbf{p},s) \gamma^0\) and
\(\overline{v}(\mathbf{p},s) 
= i v^\dagger(\mathbf{p},s) \gamma^0\) 
of the spinors (\ref{u(p,s)} \& \ref{v(p,s)}) are
\beq
\!\!\!\!\!\!\!\!\!\!\!\!\!\!\!\!\!\!\!\!\!\!\!\!\!
\overline{u}(\mathbf{p},s) = \overline{u}(\mathbf{0},s)
\frac{\left(m - i p_a \gamma^a \right)}
{\sqrt{2 p^0 (p^0+m)}} \quad \mathrm{and} \quad 
\overline{v}(\mathbf{p},s) = \overline{v}(\mathbf{0},s)
\frac{\left(m + i p_a \gamma^a \right)}
{\sqrt{2 p^0 (p^0+m)}}.
\label {barspinors}
\eeq
\par
The transposes of the spinors (\ref{u(p,s)} \& \ref{v(p,s)}) 
are by (\ref{transpose})
\beq
u^\top(\mathbf{p},s) = u^\top(\mathbf{0},s)
\frac{\gamma^0 \gamma^2 \left(m + i p_a \gamma^a \right)
\gamma^0 \gamma^2}
{\sqrt{2 p^0 (p^0+m)}}
\label {uT}
\eeq
and
\beq
v^\top(\mathbf{p},s) = v^\top(\mathbf{0},s)
\frac{\gamma^0 \gamma^2 \left(m - i p_a \gamma^a \right)
\gamma^0 \gamma^2}
{\sqrt{2 p^0 (p^0+m)}}.
\label {vT}
\eeq
\subsection{Dirac Equation\label{DE}}
These matrix formulas (\ref{u(p,s)} \& \ref{v(p,s)}) 
imply that the spinors \(u(\mathbf{p},s)\) 
and \(v(\mathbf{p},s)\) are eigenvectors of
\( -i  p_a \gamma^a \) with eigenvalues \(\pm m\):
\bea
\left(i p_a \gamma^a + m \right) u(\mathbf{p},s) & = &
\left(i p_a \gamma^a + m \right) 
\frac{\left(m - i p_b \gamma^b \right) \, u(\mathbf{0},s)}
{\sqrt{2 p^0 (p^0+m)}} \nn\\
& = &  \frac{\left(m^2 + p^ap_a \right) \, u(\mathbf{0},s)}
{\sqrt{2 p^0 (p^0+m)}} = 0
\label {momDirequ}
\eea
and
\bea
\left(m - i p_a \gamma^a \right) v(\mathbf{p},s) & = &
\left(m - i p_a \gamma^a \right) 
\frac{\left(m + i p_a \gamma^a \right) \, v(\mathbf{0},s)}
{\sqrt{2 p^0 (p^0+m)}} \nn\\
& = &  \frac{\left( m^2 + p^ap_a \right) \, v(\mathbf{0},s)}
{\sqrt{2 p^0 (p^0+m)}} = 0.
\label {momDireqv}
\eea
That is, they satisfy the Dirac equation
in momentum space.
\par
Thus the Majorana field (\ref{MajField0})
satisfies the Dirac equation
\beq
\left( \gamma^a \partial_a + m \right) \chi(x) = 0
\label {MsatD}
\eeq
in position space.
\par
The adjoint spinors (\ref{barspinors}) also satisfy
the Dirac equation in momentum space
\beq
\overline{u}(\mathbf{p},s) \left(
m + i p_a \gamma^a \right) = 0 
\quad \mbox{and} \quad
\overline{v}(\mathbf{p},s) \left(
m - i p_a \gamma^a \right) = 0 .
\label {barmomDirequv}
\eeq
\subsection{Majorana Condition}
The spinors (\ref{u(p,s)} \& \ref{v(p,s)}) satisfy 
the Majorana conditions
\beq
u(\mathbf{p},s) = \gamma^2 v^*(\mathbf{p},s) 
\quad \mbox{and} \quad
v(\mathbf{p},s) = \gamma^2 u^*(\mathbf{p},s) 
\label {MajConuv}
\eeq
as they do (\ref{MajCond0}) at 
\(\mathbf{p} = \mathbf{0} \)\@.
\subsection{Spin Sums}
The matrix formulas  (\ref{u(p,s)} \& \ref{v(p,s)}) 
for the spinors \( u(\mathbf{p},s) \) and 
\( v(\mathbf{p},s) \) simplify
the evaluation of spin sums.
We start with the spin sums over 
the \( \mathbf{p} = 0 \) spinors
(\ref{u(0)} \& \ref{v(0)}) 
\beq
\sum_s u(\mathbf{0},s) \, \overline{u}(\mathbf{0},s) = 
\half ( i \gamma^0 + I )
\label {spinsumu0}
\eeq
and
\beq
\sum_s v(\mathbf{0},s) \, \overline{v}(\mathbf{0},s) = 
\half ( i \gamma^0 -I )
\label {spinsumv0}
\eeq
as well as
\beq
\sum_s u(\mathbf{0},s) \, v^\top(\mathbf{0},s) = 
\thalf ( I + i \gamma^0 ) \gamma^2
\label {uv0}
\eeq
and
\beq
\sum_s v(\mathbf{0},s) \, u^\top(\mathbf{0},s) = 
\thalf ( I - i \gamma^0 ) \gamma^2.
\label {vu0}
\eeq
\par
Thus by 
(\ref{u(p,s)}, \ref{barspinors}, \& \ref{spinsumu0}),
the spin sum of the outer products 
\( u(\mathbf{p},s) \overline{u}(\mathbf{p},s) \) is
\beq
\sum_s u(\mathbf{p},s) \overline{u}(\mathbf{p},s)
= \frac{
\left(m - i  p_a \gamma^a \right) \, 
( i \gamma^0 + I )
\left(m - i p_b \gamma^b \right)}{4p^0(p^0+m)} 
\label {uubar1}
\eeq
which the gamma-matrix algebra (\ref{AC}) 
and the mass-shell relation \( p^2 = m^2 \)
reduce to
\beq
\sum_s u(\mathbf{p},s) \overline{u}(\mathbf{p},s)
=  \frac{ m - i p_a \gamma^a }{2p^0}.
\label {uubar}
\eeq
The spin sum with 
\( u^\dagger(\mathbf{p},s) 
= i \overline{u}(\mathbf{p},s) \gamma^0 \) is
\beq
\sum_s u(\mathbf{p},s) \, u^\dagger(\mathbf{p},s)
= \frac{( i m + p_a \gamma^a )\gamma^0} {2p^0}
\label {uudag}
\eeq
\par
Similarly by 
(\ref{v(p,s)}, \ref{barspinors}, and \ref{spinsumv0}), 
the spin sum of the outer products
\( v(\mathbf{p},s) \overline{v}(\mathbf{p},s) \) is
\beq
\sum_s  v(\mathbf{p},s) \overline{v}(\mathbf{p},s)
= \frac{\left(m + i p_a \gamma^a \right) \, 
( i \gamma^0 - I )
\left(m + i \gamma^{b} p_b \right)}
{4p^0(p^0+m)}
\label {vvbar1}
\eeq
which, 
differing from (\ref{uubar1})
as it does by \( i \to -i \) and by an overall minus sign,
is
\beq
\sum_s  v(\mathbf{p},s) \overline{v}(\mathbf{p},s)
= \frac{\mbox{} - m - i p_a \gamma^a }{2p^0}.
\label {vvbar}
\eeq
The spin sum with 
\( v^\dagger(\mathbf{p},s) 
= i \overline{v}(\mathbf{p},s) \gamma^0 \) is
\beq
\sum_s  v(\mathbf{p},s) v^\dagger(\mathbf{p},s)
 = \frac{
\left( p_a \gamma^a -i m \right) \, \gamma^0}
{2p^0}.
\label {vvdag}
\eeq
\par
With a little more effort, one finds from 
(\ref{u(p,s)}, \ref{uv0}, \& \ref{vT})
the spin sum
\bea
\sum_s 
u(\mathbf{p},s) \, v^\top(\mathbf{p},s) & = &
\frac{(m - i p_a \gamma^a ) 
(1 + i \gamma^0)\gamma^2 \gamma^0 \gamma^2
(m - i  p_b \gamma^b ) \gamma^0 \gamma^2}
{4p^0(p^0+m)} \nn\\
& = &
\frac{( im + p_a \gamma^a) \gamma^0 \gamma^2}{2p^0}.
\label {uvT}
\eea
Similarly, the \(v u^\top \) spin sum follows
from (\ref{v(p,s)}, \ref{vu0}, \& \ref{uT})
\bea
\sum_s 
v(\mathbf{p},s) \, u^\top(\mathbf{p},s) & = &
\frac{(m + i p_a \gamma^a) 
(1 - i \gamma^0) \gamma^2\gamma^0 \gamma^2
(m + i p_b \gamma^b ) \gamma^0 \gamma^2}
{4p^0(p^0+m)} \nn\\
& = & 
\frac{( \mbox{} - im + p_a \gamma^a)\gamma^0\gamma^2}{2p^0}
\label {vuT}
\eea
and so differs from (\ref{uvT}) by 
\( i \to -i \), as it must.
\subsection{Inner Products of Spinors}
The matrix formula (\ref{u(p,s)}) for 
the spinor \( u(\mathbf{p},s) \),
the behavior (\ref{HC}) of the \( \gamma \)-matrices
under hermitian conjugation, and the 
eigenvalue relation 
\( i \gamma^0 u(\mathbf{0},s) = u(\mathbf{0},s) \)
(\ref{igamma0}) imply that
\beq
u^\dagger(\mathbf{p},s) \,
u(\mathbf{p},s') = \delta_{s s'}.
\label {uDagu}
\eeq
Similarly
\beq
v^\dagger(\mathbf{p},s) \,
v(\mathbf{p},s') = \delta_{s s'}
\label {vDagv}
\eeq
and 
\( u^\dagger(\mathbf{p},s) \, v(- \mathbf{p},s') = 0 \)\@.
\par
By (\ref{igamma0}), 
the zero-momentum spinors
\( u(\mathbf{0},s) \) and \( v(\mathbf{0},s') \) 
are eigenvectors of the hermitian matrix 
\( i \gamma^0 \) with eigenvalues \(+1\) \& \(-1 \)\@.
So these spinors are orthogonal,
\( u^\dagger(\mathbf{0},s) v(\mathbf{0},s') = 0 \)\@.
Also, \( \bar u(\mathbf{0},s) u(\mathbf{0},s') = 
u^\dagger(\mathbf{0},s) i \gamma^0 u(\mathbf{0},s') 
= u^\dagger(\mathbf{0},s) u(\mathbf{0},s') =
\delta_{s s'} \) and
\( \bar v(\mathbf{0},s) v(\mathbf{0},s') = 
v^\dagger(\mathbf{0},s) i \gamma^0 v(\mathbf{0},s') 
= - v^\dagger(\mathbf{0},s) v(\mathbf{0},s') =
- \delta_{s s'} \)\@.
Now \( i \gamma^0 \vec \gamma \, u(\mathbf{0},s)
= - \vec \gamma \, i \gamma^0 u(\mathbf{0},s) =
- \vec \gamma \, u(\mathbf{0},s) \), and so the spinors
\( \vec \gamma \, u(\mathbf{0},s) \) and 
\( u(\mathbf{0},s') \)
are eigenvectors of the hermitian matrix 
\( i \gamma^0 \) with different eigenvalues
and so must be orthogonal,
\( 
u^\dagger(\mathbf{0},s) \vec \gamma \, u(\mathbf{0},s')
= 0 \)\@.  Similarly, 
\( 
v^\dagger(\mathbf{0},s) \vec \gamma \, v(\mathbf{0},s')
= 0 \)\@.  
It follows therefore from the matrix 
formulas (\ref{u(p,s)}) for \( u(\mathbf{p},s) \)
and (\ref{barspinors}) for \( \bar u(\mathbf{p},s') \)
that 
\beq
\bar u(\mathbf{p},s) \, u(\mathbf{p},s') = 
\frac{m}{p^0} \bar u(\mathbf{0},s) \, u(\mathbf{0},s')
= \frac{m}{p^0} \, \delta_{s s'} .
\label {ubaru}
\eeq
Similarly
\beq
\bar v(\mathbf{p},s) \, v(\mathbf{p},s') = 
\frac{m}{p^0} \bar v(\mathbf{0},s) \, v(\mathbf{0},s') 
= - \frac{m}{p^0} \, \delta_{s s'} .
\label {vbarv}
\eeq
\par
Since the spinors \( u \) and \( v \) 
obey the Dirac equation 
in momentum space (\ref{momDirequ} \& \ref{momDireqv}),
it follows that
\( \bar u(\mathbf{p},s) \gamma^a  
( m + i p_b' \gamma^b )
u(\mathbf{p}',s') = 0 \) and 
\( \bar u(\mathbf{p},s) ( m + i p_b \gamma^b )
\gamma^a u(\mathbf{p}',s') = 0 \)\@.
So
\beq
2m \bar u(\mathbf{p},s) \gamma^a u(\mathbf{p}',s')
= -i \bar u(\mathbf{p},s) \left(
p_b \gamma^b \gamma^a + p'_b \gamma^a \gamma^b 
\right) u(\mathbf{p}',s') 
\eeq
and since 
\(2 \gamma^b \gamma^a = \{ \gamma^b, \gamma^a \}
+ [ \gamma^b, \gamma^a ] \)
one has
\beq
\!\!\!\!\!\!\!\!\!\!\!\!\!\!\!\!\!\!\!\!\!\!
\bar u(\mathbf{p},s) \gamma^a u(\mathbf{p}',s')
= \frac{-i}{2m} \, 
\bar u(\mathbf{p},s) \left(
p^a + p^{' a} + \thalf (p_b - p'_b )
\, [ \gamma^b, \gamma^a ] \right)
u(\mathbf{p}',s') 
\label {Gordonu}
\eeq
which is Gordon's identity.
Similarly,
\beq
\!\!\!\!\!\!\!\!\!\!\!\!\!\!\!\!\!\!\!\!\!\!
\bar v(\mathbf{p},s) \gamma^a v(\mathbf{p}',s')
= \frac{i}{2m} \, 
\bar v(\mathbf{p},s) \left(
p^a + p^{' a} + \thalf (p_b - p'_b )
\, [ \gamma^b, \gamma^a ] \right)
v(\mathbf{p}',s') .
\label {Gordonv}
\eeq
\subsection{Explicit Formulas for Spinors}
The static spinors (\ref{u(0)} \& \ref{v(0)}) and
the matrix formulas (\ref{u(p,s)} \& \ref{v(p,s)})  
give explicit expressions for the spinors at 
arbitrary momentum \(\mathbf{p}\):
\beq
%\!\!\!\!\!\!\!\!\!\!\!\!\!\!\!\!\!
%\!\!\!\!\!\!\!\!\!\!\!\!\!\!\!
u({\mathbf{p},\thalf}) = \frac{1}{2\sqrt{p^0(p^0+m)}}
\pmatrix{m+p^0-p_3\cr 
         -p_1 - i p_2\cr
         m+p^0+p_3\cr 
         p_1 +ip_2\cr}
\label {u(p,+)}
\eeq
\beq
u(\mathbf{p},-{\thalf}) = \frac{1}{2\sqrt{p^0(p^0+m)}}
\pmatrix{-p_1+ip_2\cr 
         m+p^0+p_3\cr
         p_1-ip_2\cr 
         m+p^0-p_3\cr}
\label{u(p,-)}
\eeq
\beq
v({\mathbf{p},\thalf}) = \frac{1}{2\sqrt{p^0(p^0+m)}}
\pmatrix{-p_1+ip_2\cr 
         m+p^0+p_3\cr
         -p_1+ip_2\cr 
        -m-p^0+p_3\cr}
\label {v(p,+)}
\eeq
\beq
v(\mathbf{p},{-\thalf}) = \frac{1}{2\sqrt{p^0(p^0+m)}}
\pmatrix{-m-p^0+p_3\cr 
          p_1+ip_2\cr
          m+p^0+p_3\cr 
          p_1+ip_2\cr}
\label {v(p,-)}
\eeq
which,
as \(\mathbf{p} \to \mathbf{0}\), 
reduce 
to the static spinors (\ref{u(0)} \& \ref{v(0)})\@.
\subsection{Helicity}
For momenta in the z-direction
and in the limit of small
\( m / p_3 \),
these formulas yield
\beq
%\!\!\!\!\!\!\!\!\!\!\!\!\!\!\!\!\!
%\!\!\!\!\!\!\!\!\!\!\!\!\!\!\!
u({\mathbf{p},\thalf}) \approx 
\left(1 - \frac{m}{p} \right)
\pmatrix{m/(2p)\cr 
         0\cr
         1 + m/(2p)\cr 
         0\cr}
\label {u(p3,+)}
\eeq
\beq
u(\mathbf{p},-{\thalf}) \approx 
\left(1 - \frac{m}{p} \right)
\pmatrix{0\cr 
         1 + m/(2p)\cr
         0\cr 
         m/(2p)\cr}
\label {u(p3,-)}
\eeq
\beq
v({\mathbf{p},\thalf}) \approx 
\left(1 - \frac{m}{p} \right)
\pmatrix{0\cr 
         1 + m/(2p)\cr
         0\cr 
         -m/(2p)\cr}
\label {v(p3,+)}
\eeq
\beq
v(\mathbf{p},{-\thalf}) \approx 
\left(1 - \frac{m}{p} \right)
\pmatrix{-m/(2p)\cr 
          0\cr
          1 + m/(2p)\cr 
          0\cr}
\label {v(p3,-)}
\eeq
in which \( p = |\mathbf{p}| = p_3 \ge 0 \)\@.
\section{Majorana Field \label {MF}}
In this section,
we apply our spinor formulas to the Majorana field,
which is simpler and more fundamental than the Dirac field.
\par
In terms of the annihilation and creation
operators (\ref{[aa*]} \& \ref{[aa]}) and 
the spinors (\ref{u(p,s)} \& \ref{v(p,s)}),
the Majorana field is
\beq
\!\!\!\!\!\!\!\!\!\!\!\!\!\!
\chi_b(x) = \int \!\! \frac{d^3p}{(2\pi)^{3/2}}
\sum_s
\left[ 
u_b(\mathbf{p},s) \, a(\mathbf{p},s) e^{ipx} 
+     
v_b(\mathbf{p},s) \, a^\dagger(\mathbf{p},s) e^{-ipx}
\right].
\label {MajField1} 
\eeq
It satisfies the Majorana condition
\beq
\chi(x) = \gamma^2 \, \chi^*(x)
\label {Majorana condition}
\eeq
because the spinors \( u(\mathbf{p},s) \)
and \(  v(\mathbf{p},s) \) do
(\ref{MajConuv})\@.
It obeys the Dirac equation 
\beq
\left( \gamma^a \partial_a + m \right) \chi(x) = 0
\label {MajDir Eq}
\eeq
because the spinors do so
in momentum space (\ref{momDirequ} \& \ref{momDireqv}).
\par
An action density that leads to this Dirac
equation is
\beq
\mathcal{L}_M = - \tcar \bar \chi \gamma^a \partial_a \chi 
+ \tcar (\partial_a \bar \chi ) \gamma^a \chi
- \thalf m \bar \chi \chi 
\label {LM}
\eeq
in which \( - (m/2) \bar \chi \chi \)
is a Majorana mass term.
If there are several Majorana fields,
then the symmetry 
\( \bar \chi_i \chi_j = \bar \chi_j \chi_i =
(\bar \chi_i \chi_j)^\dagger \) implies
that the matrix of coefficients \( m_{ij} \)
is real and symmetric.
\subsection{Helicity}
In the limit \( m/p_3 \to 0+\),
we may infer
from the spinor formulas 
(\ref{u(p3,+)}--\ref{v(p3,-)}) and from
the form (\ref{MajField1})
of the Majorana field
that its upper two components
annihilate particles of negative helicity 
and create particles of positive helicity,
while its lower two components 
annihilate particles of positive helicity 
and create particles of negative helicity.
\par
More generally, from the explicit spinor formulas
(\ref{u(p,+)}--\ref{v(p,-)}) and from (\ref{MajField1}),
we may infer that particles created by the field 
\( (1 + \gamma_5 ) \chi \)
are partially positively polarized,
while those created by 
the field \( \chi^\dagger (1 + \gamma_5 ) \)
are partially negatively polarized.
The weak charged current selects these upper
two components.
\subsection{Causality}
The spin sums (\ref{uvT} \& \ref{vuT}) 
imply that the anti-commutator of two of 
its components is
\bea
\!\!\!\!\!\!\!\!\!\!\!\!\!\!\!\!\!
\!\!\!\!\!\!\!\!\!\!\!\!\!\!\!
\!\!\!\!\!\!\!\!
\left\{ \chi_a(x), \chi_b(y) \right\} & = & 
\int \!\! \frac{d^3p}{(2\pi)^3} \!
\sum_s 
\left( 
u_a(\mathbf{p},s) v_b(\mathbf{p},s) 
e^{ip(x-y)} 
+ 
v_a(\mathbf{p},s) u_b(\mathbf{p},s)
e^{-ip(x-y)} 
\right) \nn\\
& = & \int \!\! \frac{d^3p}{(2p^0)(2\pi)^3} \!
\left[ 
\left( (im + \gamma^c p_c)\gamma^0 \gamma^2 \right)_{ab}
e^{ip(x-y)} \right. \nn\\
& & \left. \mbox{} \qquad
+ \left( (-im +  \gamma^c p_c)\gamma^0 \gamma^2  \right)_{ab}
e^{-ip(x-y)} 
\right] \nn\\
& = & i\left( ( m - \gamma^c \partial_c )
\gamma^0 \gamma^2  \right)_{ab} \Delta(x-y)
\label {ACchichi}
\eea
in which \( \Delta(x-y) \)
is the Lorentz-invariant function
\beq
\Delta(x-y) = \int \!\! \frac{d^3p}{(2p^0)(2\pi)^3} \!
\left( e^{ip(x-y)} - e^{-ip(x-y)} \right) 
\label {Delta}
\eeq
which vanishes at space-like separations.
The equal-time anti-commutator is
\beq
\left.
\left\{ \chi_a(x), \chi_b(y) \right\}
\right|_{x^0 = y^0} = \gamma^2_{ab} \, \delta(\mathbf{x-y}).
\label {delta}
\eeq
\par
Similarly, the spin sums (\ref{uubar} \& \ref{vvbar})
imply that 
\beq
\left\{ \chi_a(x), \overline{\chi_b}(y) \right\} =
\left( m - \gamma^c \partial_c \right)_{ab}
\, \Delta(x-y)
\label {ACbar}
\eeq
so that the equal-time anti-commutator is
\beq
\left.
\left\{ \chi_a(x), \overline{\chi_b}(y) \right\} 
\right|_{x^0 = y^0} = i \gamma^0_{ab} \, \delta(\mathbf{x-y})
\label {ACchichibar}
\eeq
or equivalently
\beq
\left.
\left\{ \chi_a(x), \chi_b^\dagger(y) \right\}
\right|_{x^0 = y^0} = \delta_{ab} \, \delta(\mathbf{x-y}).
\label {ACchichidag}
\eeq
\subsection{Propagator}
In the usual way~\cite{WeinbergI}, by using the spin sums
(\ref{uubar} \& \ref{vvbar}),
one may evaluate the propagator 
\beq
\langle 0 | T\left\{ \chi_a(x) \bar \chi_b(y) \right\}
| 0 \rangle
= \int \!\!\frac{d^4p}{(2\pi)^4}
\frac{\gamma^c p_c + im}{p^2 + m^2 -i\epsilon}
\, e^{ip(x-y)}.
\label {chiProp}
\eeq
\subsection{Scalar-like Field}
The matrix formulas (\ref{u(p,s)} \& \ref{v(p,s)}) 
imply that 
we may define the Majorana field \( \chi(x) \)
in terms of the simpler field
\beq
\!\!\!\!\!\!\!\!\!\!\!\!\!\!\!\!\!\!\!\!\!\!\!\!\!\!\!\!\!\!
\!\!\!\!\!\!\!\!
\phi(x) = \int \!\! \frac{d^3p}{\sqrt{(2\pi)^3 2p^0(p^0+m)}}
\sum_{s = -{\shalf}}^{\shalf}
\left[u(\mathbf{0},s) \, a(\mathbf{p},s) e^{ipx}
+     v(\mathbf{0},s) \, a^\dagger(\mathbf{p},s) e^{-ipx}
\right] 
\label{prefieldSpinors}
\eeq
as
\beq
\chi(x) = ( m - \gamma^a \partial_a )  \phi(x).
\label{Majorana field}
\eeq
Because the \(\mathbf{p}=\mathbf{0} \) 
spinors \( u(\mathbf{0},s) \) and \(  v(\mathbf{0},s) \) 
are independent of momentum,
the field \( \phi(x) \) is like a scalar field ---
or like four scalar fields. 
\par
Since \(m^2 + p^2 = m^2 + \mathbf{p}^2 - (p^0)^2 = 0\),
the scalar-like field \(\phi(x)\) satisfies the 
Klein-Gordon equation
\beq
(m^2 + \partial_0^2 - \nabla^2) \phi(x)
= (m^2 - \eta^{ab} \partial_a \partial_b) \phi(x) = 0.
\eeq
\par
The derivative formula (\ref{Majorana field})
for the Majorana field \(\chi(x)\) implies
that it satisfies the Dirac equation:
\bea
\left( \gamma^a \partial_a + m \right) \chi(x) &=& 
\left( \gamma^a \partial_a + m \right) 
\left( m - \gamma^a \partial_a \right)  \phi(x)\cr
&=& 
\left( m^2 - \gamma^a \gamma^b \partial_a \partial_b \right) 
\phi(x)\cr 
&=& \left( m^2 - \thalf (\{ \gamma^a, \gamma^b \} +
[ \gamma^a, \gamma^b ] )
\partial_a \partial_b \right)  \phi(x) \cr 
&=&  
\left( m^2 - \eta^{ab}\partial_a \partial_b \right) \phi(x) 
= 0.
\label{MajDirac equation} 
\eea
\par
The adjoint Majorana field 
\(\overline \chi(x) = i \chi^\dagger(x) \gamma^0\) is
\beq
\overline \chi(x) = \overline \phi(x) \, 
\left( m + \gamma^a \stackrel{\leftarrow}{\partial_a} \right)
\label {phibar}
\eeq
in which the derivatives act to the left.
\section{Dirac Field \label {DF}}
In this section,
we construct a Dirac field 
from two Majorana fields 
that describe particles of the same mass.
\par
Suppose there are two spin-one-half particles
of the same mass \(m\) described by the
two operators \(a_1(\mathbf{p},s)\) and
\(a_2(\mathbf{p},s)\) which satisfy
the anti-commutation relations
\beq
\{ a_i(\mathbf{p},s),  a_j^\dagger(\mathbf{p'},s') \}
= \delta_{i j} \, 
\delta_{s s'} \delta^3(\mathbf{p} - \mathbf{p'}).
\eeq
Then we have two Majorana fields
\beq
\!\!\!\!\!\!\!\!\!\!\!\!\!\!
\chi_{bi}(x) = \int \!\! \frac{d^3p}{(2\pi)^{3/2}}
\sum_s
\left[ 
u_b(\mathbf{p},s) \, a_i(\mathbf{p},s) e^{ipx} 
+     
v_b(\mathbf{p},s) \, a_i^\dagger(\mathbf{p},s) e^{-ipx}
\right]
\label{MajField} 
\eeq
for \(i = 1, 2\)
that satisfy the same Dirac equation
\beq
\left( \gamma^a \partial_a + m\right) \, \chi_i(x) = 0.
\label {diraci}
\eeq
So it makes sense to combine them into one Dirac
field
\beq
\psi(x) = \frac{1}{\sqrt{2}} 
\left[ \chi_1(x) + i \chi_2(x) \right] 
\label{DiracField}
\eeq
which satisfies the Dirac equation
\beq
\left( \gamma^a \partial_a + m \right) \, \psi(x) = 0
\label {diraceq}
\eeq
because its Majorana parts do.
In terms of the same spinors (\ref{u(p,s)} \& \ref{v(p,s)}),
the Dirac field is 
\beq
\!\!\!\!\!\!\!\!\!\!\!\!\!\!\!\!\!\!\!\!\!\!\!\!\!
\psi(x) = \int \frac{d^3p}{(2\pi)^{3/2}}
\sum_s
\left[ 
u(\mathbf{p},s) \, a(\mathbf{p},s) e^{ipx}
+     
v(\mathbf{p},s) \, a^{c \, \dagger}(\mathbf{p},s) e^{-ipx}
\right]
\label {DiracFieldSpinors} 
\eeq
with the complex operators
\beq
a(\mathbf{p},s) = \frac{1}{\sqrt{2}} 
\left[a_1(\mathbf{p},s) + i a_2(\mathbf{p},s) \right]
\label{complex a}
\eeq
and
\beq
a^{c\dagger}(\mathbf{p},s) = \frac{1}{\sqrt{2}} 
\left[a_1^\dagger(\mathbf{p},s) + 
i a_2^\dagger(\mathbf{p},s) \right],
\label{complex acdag}
\eeq 
whence
\beq
a^c(\mathbf{p},s) = \frac{1}{\sqrt{2}} 
\left[a_1(\mathbf{p},s) - i a_2(\mathbf{p},s) \right].
\label{complex ac}
\eeq 
These complex annihilation and creation operators
satisfy the anti-commutation relations
\beq
\{ a(\mathbf{p},s),  a(\mathbf{p}',s') \} = 0
\qquad
\{ a^c(\mathbf{p},s),  a^c(\mathbf{p}',s') \} = 0
\label {aa+acac+}
\eeq
and
\beq
\{ a(\mathbf{p},s),  a^c(\mathbf{p}',s') \} = 0
\qquad
\{ a(\mathbf{p},s),  
a^{c \dagger}(\mathbf{p}',s') \} = 0
\label {aac+aa*c+}
\eeq
as well as
\beq
\{ a(\mathbf{p},s),  a^\dagger(\mathbf{p}',s') \} = 
\delta_{s s'} \, \delta( \mathbf{p} - \mathbf{p}' ).
\label {aa*+}
\eeq
and
\beq
\{ a^c(\mathbf{p},s),  a^{c \dagger}(\mathbf{p}',s') \} 
= \delta_{s s'} \, \delta( \mathbf{p} - \mathbf{p}' ).
\label {acac*+}
\eeq
\par
An action density that leads to the Dirac
equation is the sum of two Majorana action densities
(\ref{LM}) of the same mass \(m\)
\beq
\mathcal{L}_D = \mathcal{L}_{M1} + \mathcal{L}_{M2}
=  - \thalf \bar \psi \gamma^a \partial_a \psi 
+ \thalf (\partial_a \bar \psi ) \gamma^a \psi
- m \bar \psi \psi .
\label {LD}
\eeq
\subsection{Anti-particle Field}
Because the Majorana components \(\chi_1\) and 
\(\chi_2\) of the Dirac field \(\psi\) satisfy
the Majorana condition (\ref{Majorana condition}),
the complex conjugate field 
\(\psi^*_a(x) = \psi^\dagger_a(x)\)
multiplied by \(\gamma^2\) is the
field of the anti-particle
\bea
\psi^c(x) & = & \gamma^2 \psi^*(x) 
= \gamma^2 \frac{1}{\sqrt{2}} 
\left[ \chi_1^*(x) - i \chi_2^*(x) \right] \nn\\
& = & \frac{1}{\sqrt{2}} 
\left[ \chi_1(x) - i \chi_2(x) \right].
\label {antifield1} 
\eea
More explicitly, the spinors
\(u(\mathbf{p},s)\) \& \( v(\mathbf{p},s) \)
satisfy the Majorana Condition
(\ref{MajConuv}) according to which they are interchanged
by the operation \(*\) followed by \(\gamma^2\),
and so these operations turn \(\psi\) into 
the charge-conjugate field \(\psi_c = \gamma^2 \psi^*\)
\bea
\!\!\!\!\!\!\!\!\!\!
\psi_c(x) & = & \int \!\! \frac{d^3p}{(2\pi)^{3/2}}
\sum_s
\left[ 
v(\mathbf{p},s) \, a^\dagger(\mathbf{p},s) e^{-ipx}
+     
u(\mathbf{p},s) \, a^{c}(\mathbf{p},s) e^{ipx}
\right] \nn\\
& = & \int \!\! \frac{d^3p}{(2\pi)^{3/2}}
\sum_s
\left[ 
u(\mathbf{p},s) \, a^{c}(\mathbf{p},s) e^{ipx}
+
v(\mathbf{p},s) \, a^\dagger(\mathbf{p},s) e^{-ipx}
\right] 
\label {charge-conjugate field}
\eea
which is the field of the anti-particle.
\subsection{Helicity}
In the limit of small, positive \( m/p_3 \),
we may infer
from the spinor formulas 
(\ref{u(p3,+)}--\ref{v(p3,-)}) and from
the form (\ref{DiracFieldSpinors})
of the Dirac field \( \psi \)
that its upper two components
annihilate particles mainly of negative helicity,
because of the coefficients \(u_1(\mathbf{p},s)\)
and \(u_2(\mathbf{p},s)\),
and create anti-particles mainly of positive helicity,
because of the coefficients \(v_1(\mathbf{p},s)\)
and \(v_2(\mathbf{p},s)\)\@.
Similarly, the upper two components of the
adjoint field \( \psi^\dagger \) 
create particles mainly of negative helicity,
because of the coefficients \(u_1^*(\mathbf{p},s)\)
and \(u_2^*(\mathbf{p},s)\),
and annihilate anti-particles mainly of positive helicity,
because of the coefficients \(v_1^*(\mathbf{p},s)\)
and \(v_2^*(\mathbf{p},s)\)\@.
\par
The factor  \( 1 + \gamma_5 \)
in the charged-current of the weak interaction
selects the upper two components.
Thus in the decay 
\( \mu^- \to \nu_\mu + e^- + \bar \nu_e \) 
of the muon, the field \( \psi_e^\dagger \) 
creates an electron of mainly negative helicity.
The electrons from unpolarized muons
mainly have negative helicity.  
Similarly, in the decay 
\( \mu^+ \to \bar \nu_\mu + e^+ + \nu_e \) 
of the positive muon, the field \( \psi_e \)
creates a positron of mainly positive helicity.
The \( \bar \nu_e \) and \( \nu_e \)
respectively and overwhelmingly
have positive and negative helicity.
\par
For the same reason, electrons emitted in
beta decay \( n \to p + e + \bar \nu_e \)
tend to be of negative helicity,
although the factor \( m/|\mathbf{p}_e| \)
need not be tiny.  The \( \bar \nu_e \) 
has positive helicity.
\par
The decay of the charged pion provides
another example.  The \(\pi^-\) can decay
into \(e + \bar \nu_e\) and into \(\mu + \bar \nu_\mu\)\@.
Since the electron is some 200 times 
lighter than the muon, the available
phase space of the \(e + \bar \nu_e\) 
channel is 3.49 times greater than that of
the \(\mu + \bar \nu_\mu\) channel.
So one would expect that the dominant decay channel 
would be \(\pi^- \to e + \bar \nu_e\)\@.
Experimentally, however, 99.9877\% of the decays go
via the channel \(\pi^- \to \mu + \bar \nu_\mu\)\@.
Why?  Well, the field \(\psi_e^\dagger\)
makes an electron of mainly negative helicity,
and the field \(\nu_e\) creates a \(\bar \nu_e\) 
of mainly positive helicity.  Now, in the rest frame
of the decaying pion, the momenta of the
electron and neutrino are (equal and) opposite,
and so in a final state composed
of the large spinor components,
their two parallel spins would add to an
angular momentum of \(\hbar\)\@.
But the pion is a pseudo-scalar meson, and so
conservation of angular momentum allows
only the small \( m_e/(2p_e) \) components
of the spinors (\ref{u(p3,+)}--\ref{v(p3,-)}) 
to contribute to the amplitude.
This effect also slows down the principal
decay mode \( \pi^- \to \mu + \bar \nu_\mu\),
but the factor \( m_\mu/(2p_\mu) \)
is bigger because \( m_\mu \approx 207 m_e \)
and because \( p_e \approx 2.34 p_\mu \)\@.
The \(e + \bar \nu_e\) channel is
helicity suppressed
by the factor 
\( [p_\mu m_e/(p_e m_\mu)]^2 = 4.2 \times 10^{-6} \)\@.
\subsection{Causality}
The Dirac field anti-commutes with itself 
\beq
\{ \psi_a(x), \psi_b(y) \} = 0
\label {psipsi0}
\eeq
because 
of the way it is constructed 
\(\sqrt{2} \, \psi = \chi_1 + i \chi_2\)
from two fields of the same mass
\beq
2 \, \{ \psi_a(x), \psi_b(y) \} = 
\{ \chi_{1a}(x), \chi_{1b}(y) \}
- \{ \chi_{2a}(x), \chi_{2b}(y) \}  = 0 .
\label {acpsipsi}
\eeq
The vanishing of this anti-commutator also
follows from the anti-commutation relations
(\ref{aa+acac+}--\ref{acac*+}).
\par
Similarly,
the anti-commutator of \(\psi(x)\) with its adjoint 
\(\overline{\psi}(x) = i\psi^\dagger(x) \gamma^0 \)
follows from the anti-commutation relation (\ref{ACbar})
obeyed by its constituent Majorana fields
\bea
\{ \psi_a(x), \overline \psi_b(y) \} & = &
\thalf 
\left\{ \chi_{1a}(x), \overline{\chi_{1b}}(y) \right\}
+ \thalf \left\{ \chi_{2a}(x), \overline{\chi_{2b}}(y) \right\} \nn\\
& = &
\left( m - \gamma^c \partial_c \right)_{ab}
\, \Delta(x-y)
\label {nicepsipsibar}
\eea
in which \( \Delta(x-y) \)
is the Lorentz-invariant function (\ref{Delta})\@.
One also may evaluate this anti-commutator by using
the anti-commutation relations
(\ref{aa+acac+} -- \ref{acac*+}) and the
spin sums (\ref{uubar} \& \ref{vvbar})
\bea
\!\!\!\!\!\!\!\!\!\!\!\!\!\!\!\!\!\!\!\!\!\!\!\!\!\!\!\!\!\!\!\!
\!\!\!\!\!\!\!\!
\{ \psi_a(x), \overline \psi_b(y) \} & = &
\int \!\! \frac{d^3p}{(2\pi)^3} \!
\sum_s 
\left( 
u_a(\mathbf{p},s) \overline{u}_b(\mathbf{p},s) 
e^{ip(x-y)} 
+ 
v_a(\mathbf{p},s) \overline{v}_b(\mathbf{p},s)
e^{-ip(x-y)} 
\right) \nn\\
& = & \int \!\! \frac{d^3p}{(2p^0)(2\pi)^3} \!
\left( 
\left( m - i \gamma^c p_c \right)_{ab}
e^{ip(x-y)}  
- \left( m + i \gamma^c p_c \right)_{ab}
e^{-ip(x-y)} 
\right) \nn\\
& = & ( m - \gamma^c \partial_c )_{ab} \Delta(x-y).
\label {ACpsipsibar}
\eea
At equal times, \( \Delta(x-y) \) vanishes,
and only the time derivative contributes in (\ref{ACpsipsibar})
\beq
\left.
\{ \psi_a(x), \overline \psi_b(y) \} \right|_{x^0=y^0} =
i \gamma^0_{ab} \, \delta(\mathbf{x-y})
\label {ETACpsipsibar}
\eeq
or 
\beq
\left.
\{ \psi_a(x), \psi^\dagger_b(y) \} \right|_{x^0=y^0} = 
\delta_{ab} \, \delta(\mathbf{x-y}).
\label {ETACpsipsi*}
\eeq
\subsection{Propagator}
Since the Dirac field is the combination (\ref{DiracField})
of Majorana fields,
its propagator 
\( \langle 0 | T\{ \psi_a(x) \bar \psi_b(y) | 0 \rangle \)
is the same as that of a Majorana field 
(\ref{chiProp}) of the same mass.
\subsection{Complex Scalar-like Field}
\par
The Dirac field also may be obtained from 
the scalar-like field 
\beq
\!\!\!\!\!\!\!\!\!\!\!\!\!\!\!\!\!\!
\!\!\!\!\!\!\!\!\!\!\!\!\!
\!\!\!\!\!\!\!
\Phi(x) = \int \!\! \frac{d^3p}{\sqrt{2(2\pi)^3 p^0(p^0+m)}}
\!\sum_{s = -{\shalf}}^{\shalf}
\left[u(\mathbf{0},s) a(\mathbf{p},s) e^{ipx}
+     v(\mathbf{0},s) a^{c \dagger}(\mathbf{p},s) e^{-ipx}
\right] 
\label{CprefieldSpinors}
\eeq
as
\beq
\psi(x) = \left( m - \gamma^a \partial_a \right)
 \Phi(x) .
\label{DiracFieldPhi}
\eeq
The field \(\Phi(x)\) is like a scalar field 
because the static spinors
(\ref{u(0)} \& \ref{v(0)})
do not vary with the momentum.
It satisfies anti-commutation relations:
\beq
\{ \Phi_a(x), \Phi_b(y) \} = 0.
\label {PhiAC}
\eeq
With this construction, the Dirac field satisfies 
the Dirac equation
\beq
\left( \gamma^a \partial_a + m \right) \psi(x) = 0
\eeq
because the field \( \Phi(x) \)
satisfies the Klein-Gordon equation,
and it satisfies the anti-commutation relations
\beq
\{ \psi_a(x), \psi_b(y) \} = 0
\label {psiAC}
\eeq
because the complex scalar-like field \(\Phi(x)\) does,
Eq.(\ref{PhiAC})\@.
The adjoint Dirac field 
\( \overline \psi (x) = i \psi^\dagger (x) \gamma^0 \) 
is related to the scalar-like field \( \Phi(x) \) by
\beq
\overline \psi(x) = \overline \Phi(x) \, 
\left( m + 
\gamma^a \stackrel{\leftarrow}{\partial_a} \right).
\label {psibar}
\eeq
\section{Applications to Supersymmetry\label{susy}}
The relations we have derived for Majorana
and Dirac fields are useful in many contexts. 
As an example of their utility, 
we shall in this final section use some of them
to determine the
generators of supersymmetry in the Wess-Zumino model.
Such supercharges play key roles in supersymmetric
field theories.
\subsection{Wess-Zumino Model}
If \( \chi \) is a Majorana field,
\(\chi = \gamma^2 \chi^* \), \( B \) a real scalar field, 
and \( C \) a real pseudo-scalar field, then 
the action density of the 
Wess-Zumino model~\cite{Wess1974} is
\bea
\mathcal{L} & = & - \thalf \partial_a B \, \partial^a B 
- \thalf \partial_a C \, \partial^a C 
- \tcar \bar \chi \gamma^a \partial_a \chi 
+ \tcar (\partial_a \bar \chi ) \gamma^a \chi \nn\\
& & + \thalf \left( F^2 + G^2 \right)
+ m \left( FB + GC - \thalf \bar \chi \chi \right) 
\nn\\
& & + g \left[
F \left( B^2 - C^2 \right) + 2 G B C 
- \bar \chi \left( B + i \gamma_5 C \right) \chi
\right] 
\label {WZL}
\eea
in which \( (-m/2) \, \bar \chi \chi \)
is a Majorana mass term.
The action density \( \mathcal{L} \)
is said to be supersymmetric
because it changes only by a total divergence
under the susy transformation
\bea
\delta B & = & \bar \chi \, \alpha \qquad 
\delta C = - i \bar \chi \, \gamma_5 \alpha \nn\\
\delta \chi & = & \partial_a \left( 
B + i \gamma_5 C \right) \gamma^a \alpha
+ \left( F - i \gamma_5 G \right) \alpha \nn\\
\delta F  & = & - \partial_a \bar \chi \, \gamma^a \alpha 
\qquad \delta G = 
i \partial_a \bar \chi \gamma^a \gamma_5 \, \alpha
\label {susyWZ}
\eea
in which \( \alpha \) is a constant anti-commuting 
c-number spinor that satisfies the Majorana condition
\(\alpha = \gamma^2 \alpha^* \)\@.
The change \( \delta \mathcal{L} \)
is a total divergence irrespective
of whether the fields obey their
of the equations of motion.
Some authors write the action density (\ref{WZL})
and the susy transformation (\ref{susyWZ}) in terms
of \(C' = -C\) and \(\delta C' = -\delta C\)\@. 
\par
We have written the susy transformation (\ref{susyWZ})
exclusively in terms of the spinor \(\alpha\) by using
the Majorana character of \(\chi\) and of \(\alpha\)
which imply 
\beq
\bar \alpha \chi = \bar \chi \alpha
\quad \mbox{and} \quad 
\bar \alpha \gamma_5 \chi = \bar \chi \gamma_5 \alpha.
\label {albarchi}
\eeq
Since \( \gamma^2 (\gamma^a)^* = \gamma^a \gamma^2 \),
it follows from (\ref{susyWZ}) that 
the change \(\delta \chi\)
is also Majorana
\beq
\delta \chi = \gamma^2 \chi^* 
\label {dchiM}
\eeq
which in turn implies both
\beq
\delta \bar \chi \, \chi = \bar \chi \, \delta \chi
\quad \mbox{and} \quad
\delta \bar \chi \, \gamma_5 \chi 
= \bar \chi \, \gamma_5 \delta \chi
\label {dchibarchi} 
\eeq
and with 
\( \gamma^2 (\gamma^a)^\top \gamma^0 
= \gamma^0 \gamma^a \gamma^2 \)
also
\beq
\delta \bar \chi \gamma^a \chi 
= - \bar \chi \gamma^a \delta \chi.
\label {dchibargchi}
\eeq
Incidentally, \( \overline{\delta \chi} 
= \delta \bar \chi \)\@.
\subsection{Change in Action Density}
Under the susy transformation (\ref{susyWZ}),
the change in the action density is
\bea
\delta \mathcal{L} & = & 
\left\{ - \partial_a B \partial^a \bar \chi 
+ i \partial_a C \partial^a \bar \chi \gamma_5  \right. \nn\\
& & - \thalf \bar \chi \gamma^a \partial_a 
\left[ \partial_b \left( 
B + i \gamma_5 C \right) \gamma^b 
+ \left( F - i \gamma_5 G \right)  \right] \nn\\
& & + \thalf \partial_a \bar \chi \gamma^a 
\left[ \partial_b \left( 
B + i \gamma_5 C \right) \gamma^b 
+ \left( F - i \gamma_5 G \right)  \right] \nn\\
& & - F \partial_a \bar \chi \, \gamma^a  
+ i G \partial_a \bar \chi \gamma^a \gamma_5 \, 
+ m \left\{ F \bar \chi \,  
- B  \partial_a \bar \chi \, \gamma^a 
-i G \bar \chi \, \gamma_5  \right. \nn\\
& & \left. + 
i C  \partial_a \bar \chi \gamma^a \gamma_5 \,  
- \bar \chi \left[ \partial_a \left( 
B + i \gamma_5 C \right) \gamma^a 
+ \left( F - i \gamma_5 G \right)  
\right] \right\}  \nn\\
& & - g B^2 \partial_a \bar \chi \gamma^a 
+ g C^2 \partial_a \bar \chi \gamma^a 
+ 2g FB \bar \chi + 2igFC \bar \chi \gamma_5 \nn\\
& & + 2igBC \partial_a \bar \chi \gamma^a  \gamma_5 
- 2i GB \bar \chi \gamma_5
+ 2g GC \bar \chi \nn\\
& & - 2 g \bar \chi B \left[ \partial_a 
\left( B + i \gamma_5 C \right) \gamma^a + F - i \gamma_5 G
\right] \nn\\
& & - 2 i g \bar \chi \gamma_5 C \left[
\partial_a 
\left( B + i \gamma_5 C \right) \gamma^a + F - i \gamma_5 G
\right] \nn\\
& & \left. - g \bar \chi \chi \, \bar \chi 
- g \bar \chi \gamma_5 \chi \, \bar \chi \gamma_5 
\right\} \alpha.
\label {dLK1}
\eea
After several cancellations, \( \delta \mathcal{L} \)
simplifies to
\bea
\delta \mathcal{L} & = & 
\left\{ - \partial_a B \partial^a \bar \chi 
+ i \partial_a C \partial^a \bar \chi \gamma_5  
- \thalf \bar \chi \gamma^a  \gamma^b \partial_a 
\partial_b \left( 
B - i \gamma_5 C \right) \right. \nn\\
& & + \thalf \partial_a \bar \chi \gamma^a \gamma^b 
\partial_b \left( 
B - i \gamma_5 C \right) 
 - \thalf \partial_a \left( F \bar \chi \, \gamma^a \right)
+ \thalf i \partial_a ( G \bar \chi \gamma^a \gamma_5 ) \nn\\
& & + m \left[ - \partial_a ( B \bar \chi \, \gamma^a ) 
+ i \partial_a ( C \bar \chi \gamma^a \gamma_5 ) \right] \nn\\
& & \left. - g \left[ 
\partial_a ( \bar \chi (B + i \gamma_5 C)^2 \gamma^a )
+ \bar \chi \chi \, \bar \chi 
+ \bar \chi \gamma_5 \chi \, \bar \chi \gamma_5 \right]
\right\} \alpha.
\label {dLK2}
\eea
\par
The terms that are cubic in \( \chi \) cancel.
To see this, we note that the Majorana condition
(\ref{Majorana condition}) 
\( \chi = \gamma^2 \, \chi^* \)
and \( (\gamma^2)^2 = 1 \) imply that
\( \chi^* = \gamma^2 \chi \) and so that
\( \chi^\dagger = \chi^\top (\gamma^2)^\top 
= \chi^\top \gamma^2 \), whence
\( \bar \chi = i \chi^\dagger \gamma^0
= i \chi^\top \gamma^2 \gamma^0 \)\@.
It follows that
\beq
- \bar \chi \chi \,\, \bar \chi \alpha
= \chi^\top \pmatrix{\sigma_2 & 0 \cr
                     0 & - \sigma_2} \chi \,\,
 \chi^\top \pmatrix{\sigma_2 & 0 \cr
                     0 & - \sigma_2} \alpha 
\label {chisigma1}
\eeq
and
\beq
- \bar \chi \gamma_5 \chi \,\, \bar \chi \gamma_5 \alpha
= \chi^\top \pmatrix{\sigma_2 & 0 \cr
                     0 & - \sigma_2} \chi \,\,
 \chi^\top \pmatrix{\sigma_2 & 0 \cr
                     0 & \sigma_2} \alpha .
\label {chisigma2}
\eeq
The sum of these two terms contains only products
\(\chi_i \chi_j \chi_k \) in which all the indices
are either 1 or 2 or all are 3 or 4;
all products with three different indices,
like \( \chi_1 \chi_2 \chi_3 \),
cancel.  So every surviving
term contains a product 
of two identical fields \( \chi_i (x) \, \chi_i (x) \)
for some \( i \)\@.
But the equal-time
anti-commutation relation (\ref{delta})
(to wit,
\( \{  \chi_a(x), \chi_b(y) \}
=  \gamma^2_{ab} \, \delta(\mathbf{x-y}) \))
implies that such terms vanish.  
So the terms with three \( \chi \)'s cancel.
\par
Next, the anti-commutation relations (\ref{AC}) 
of the gamma matrices imply that
\beq
\gamma^a \gamma^b \partial_a \partial_b = 
\gamma^b \gamma^a \partial_a \partial_b = 
\left( - \gamma^a \gamma^b + 2 \eta^{ab} \right)
\partial_a \partial_b 
\label {cIid}
\eeq
so we can write \( \delta \mathcal{L} \) as
the total divergence 
\beq
\delta \mathcal{L} = \partial_a K^a 
\label {totDiv}
\eeq
of the current
\bea
K^a & = & \left[ - \bar \chi \partial^a ( B + i \gamma_5 C ) + 
\thalf \bar \chi \gamma^a \gamma^b \partial_b ( B - i \gamma_5 C )
\right. \nn\\
& & - \thalf \bar \chi \gamma^a ( F -i \gamma_5 G )
- m \bar \chi \gamma^a ( B - i \gamma_5 C ) \nn\\
& & \left. - g \bar \chi (B + i \gamma_5 C)^2 \gamma^a 
\right] \alpha .
\label {Ka}
\eea
Thus whether or not the fields satisfy their 
equations of motion, the change in the
Wess-Zumino action density is a
total divergence, and in this sense, 
the Wess-Zumino action is supersymmetric.
\subsection{Noether Current}
For a general action density \( \mathcal{L}(\phi) \),
Lagrange's equations 
\beq
\partial_a 
\frac{\partial \mathcal{L}(\phi)}{\partial \partial_a \phi_i}
= \frac{\partial \mathcal{L}(\phi)}{\partial \phi_i} 
\label {Leq}
\eeq
and the identity  
\( \delta \partial_a \phi_i = \partial_a \delta \phi_i \)
imply that any first-order change 
\beq
\delta \mathcal{L}(\phi) = 
\frac{\partial \mathcal{L}(\phi)}{\partial \partial_a \phi_i}
\delta \partial_a \phi_i + 
\frac{\partial \mathcal{L}(\phi)}{\partial \phi_i}
\delta \phi_i 
\label {dL}
\eeq
is a total divergence
\beq
\delta \mathcal{L}(\phi) = \partial_a \left(
\frac{\partial \mathcal{L}(\phi)}{\partial \partial_a \phi_i}
\delta \phi_i \right) = \partial_a J^a 
\label {totdiv}
\eeq
of a Noether current
\beq
J^a = 
\frac{\partial \mathcal{L}(\phi)}{\partial \partial_a \phi_i}
\delta \phi_i .
\label {JN}
\eeq
The Noether current \( J^a \) is conserved
\( \partial_a J^a = 0 \) 
by the equations of motion (\ref{Leq}), if 
the action density is invariant
\( \delta \mathcal{L} = 0 \) to first order. 
The Noether current of the supersymmetry 
transformation (\ref{susyWZ}) is not 
conserved.
\subsection{Wess-Zumino Noether Current}
The change in the Wess-Zumino action density
(\ref{WZL}) is by 
(\ref{dchibargchi}, 
\ref{totdiv}, \& \ref{JN})
the divergence
\beq
\delta \mathcal{L} = \partial_a J^a
\label {dLdivJ}
\eeq
of the susy Noether current 
\bea
J^a & = & \tcar \delta \bar \chi \, \gamma^a \chi
- \tcar \bar \chi \gamma^a \, \delta \chi 
-  ( \partial^a B )\, \delta B  - ( \partial^a C ) \, \delta C \nn\\
& = & - \thalf \bar \chi \gamma^a \, \delta \chi 
-  ( \partial^a B )\, \delta B  - ( \partial^a C ) \, \delta C 
\label {Ja1}
\eea
or
\bea
J^a & = & \left[ - \thalf \bar \chi \gamma^a  \gamma^b 
\partial_b \left( B - i \gamma_5 C \right) 
- \thalf \bar \chi \gamma^a 
\left( F - i \gamma_5 G \right) \right. \nn\\
& & \left.  
- \bar \chi \partial^a ( B - i \gamma_5 C )
\right] \alpha .
\label {WZSNC}
\eea
The current \( J^a \) is hermitian 
\( J_a^\dagger = J_a \) as it should be, but
because the change \( \delta \mathcal{L} \) is a non-zero
total divergence, it is not conserved.
\subsection{The Susy Current}
Although neither the current \(J^a\) nor  the current \(K^a\)
is conserved, by (\ref{totDiv} \& \ref{totdiv}),
the divergence of each of them is
the change
\( \delta \mathcal{L} \) in the action density
\beq
\partial_a K^a = \partial_a J^a =  \delta \mathcal{L}.
\label {divK-divJ}
\eeq
So the difference of the two currents
\bea
S^a & = & K^a - J^a \nn\\
& = & \bar \chi \gamma^a  \left[ \gamma^b 
\partial_b - m - g \left( B - i \gamma_5 C \right) \right]
( B - i \gamma_5 C ) \alpha 
\label {Sa}
\eea
has zero divergence
\beq
\partial_a S^a = 0
\label {divS0}
\eeq
and is conserved.
This current \( S^a \)
is the conserved susy current
of the Wess-Zumino action.
It contains no auxiliary fields.
\subsection{Supercharges}
The supercharge \( \bar Q \) 
multiplied by the spinor \( \alpha \)
is the spatial integral of \( S^0 \)
\bea
\bar Q \, \alpha & = &
\int \!\! d^3x \, S^0 \nn\\
& = & \int \!\! d^3x \, 
\bar \chi \gamma^0  \left[ \gamma^b 
\partial_b - m - g \left( B - i \gamma_5 C \right) \right]
( B - i \gamma_5 C ) \alpha \nn\\
& = & - i \int \!\! d^3x \, 
\chi^\dagger \left[ \gamma^b 
\partial_b - m - g \left( B - i \gamma_5 C \right) \right]
( B - i \gamma_5 C ) \alpha .
\label {barQa}
\eea
By inserting unity in the form \( I = (\gamma^2)^2 \),
one may show that the supercharge
satisfies the Majorana condition
\beq
Q = \gamma^2 \, Q^*
\label {Qmaj}
\eeq
so that
\bea
\!\!\!\!\!
\bar Q \alpha & = & \bar \alpha Q \\
& = & \bar \alpha \!
\int \!\! d^3x 
\left\{ \gamma^a \partial_a 
\left( B + i \gamma_5 C \right)
+ \left[ m + g \left( B - i \gamma_5 C \right)
\right] \left( B - i \gamma_5 C \right)
\right\}
\gamma^0 \chi . \nn
\label {baraQ}
\eea
\par
The susy transformation rules (\ref{susyWZ})
may be written as 
\beq
i \delta \mathcal{O}(x) = 
[ \bar Q \alpha, \mathcal{O}(x) ]
= [ \bar \alpha Q ,  \mathcal{O}(x) ] .
\label {[Qa,O]}
\eeq
The change in the field \( B(x) \) is
\bea
i \delta B(x) & = & [ \bar Q \alpha, B(x) ] \nn\\
& = & -i \int \!\! d^3x' \, 
\chi^\dagger \gamma^0 [ \partial_0 B(x'), B(x) ] 
\alpha 
\label {dB1}
\eea
which the equal-time commutation relation
\(
[ B(x) , \partial_0 B(x') ] 
= i \delta(\mathbf{x-x'})
\)
reduces to
\beq
i \delta B(x) = -i \int \!\! d^3x' \, 
\chi^\dagger \gamma^0 (-i) \delta(\mathbf{x-x'})
= i \bar \chi \alpha
\label {dB}
\eeq
in agreement with (\ref{susyWZ})\@.
Similarly, the change in \( C(x) \) is
\bea
i \delta C(x) & = & [ \bar Q \alpha, C(x) ] \nn\\
& = &  -i \int \!\! d^3x' \, 
\chi^\dagger \gamma^0 (-i \gamma_5) 
[ \partial_0 C(x'), C(x) ] \alpha =
\bar \chi \gamma_5 \alpha
\eea
as in (\ref{susyWZ})\@.
\par
By (\ref{[Qa,O]}), the change in \(\chi\) is
\bea
\!\!\!\!\!\!\!\!\!\!\!\!\!\!\!\!\!\!
i \delta \chi_a(x) & = & [ \bar Q \alpha, \chi_a(x) ] \\
& = & i \int \!\! d^3x' \, 
\{ \chi^\dagger(x'), \chi_a(x) \}
\left[ \gamma^b 
\partial_b - m - g \left( B - i \gamma_5 C \right) \right]
( B - i \gamma_5 C ) \alpha \nn
\eea
and so since
\( \{\chi_b^\dagger(x'), \chi_a (x)\}
= \delta_{ab} \delta(\mathbf{x-x'}) \)
(the equal-time anti-commutation relation 
\ref{ACchichidag}),
\(\delta \chi \) is
\beq
\delta \chi = 
\left[ \gamma^b 
\partial_b - m - g \left( B - i \gamma_5 C \right) \right]
( B - i \gamma_5 C ) \alpha .
\label {dchi1}
\eeq
\par
The auxiliary fields \( F \) and \( G \)
occur quadratically and
without their derivatives in the action density
(\ref{WZL}); their field equations are 
\beq
F = - m B - g(B^2 - C^2)
\label {F=}
\eeq
and
\beq
G = - m C - 2 g B C.
\label {G=}
\eeq
In terms of them,
the change in \(\chi\) is
\beq
\delta \chi = \partial_a \left( 
B + i \gamma_5 C \right) \gamma^a \alpha
+ \left( F - i \gamma_5 G \right) \alpha 
\label {dchi}
\eeq
as in (\ref{susyWZ})\@.
\par
The supercharges (\ref{barQa}\&\ref{baraQ})
obey the anti-commutation relation
\beq
\{ Q_a, \bar Q_b \} = 
- 2 i P_c \gamma^c_{ab} 
\label {FAC}
\eeq
which is a fundamental property of
the algebra of supersymmetric theories
with a single Majorana supercharge.
\par
The supercharges \(Q_f\) of the free theory
are given by (\ref{barQa}) or (\ref{baraQ})
with \( g = 0 \)\@.
Because the spinors \( u(\mathbf{p},s) \)
and \( v(\mathbf{p},s) \) satisfy the
Dirac equation in momentum space
(\ref{momDirequ} \& \ref{momDireqv}),
one may write \(Q_f\) as
\bea
Q_f & = &
i \! \int \!\! d^3p
\sqrt{2p^0}
\sum_{s = -{\shalf}}^{\shalf}
\left[
\left( b(p) - i \gamma_5 c(p) \right) 
v(\mathbf{p},s) a^\dagger(\mathbf{p},s) 
\right.
\nn\\
& & \qquad\qquad\quad\;\;\;
\left.
- \left(b^\dagger(p) - i \gamma_5 c^\dagger(p) 
\right) 
u(\mathbf{p},s) a(\mathbf{p},s)
\right]
\label {Qf}
\eea
from which it is clear that they
annihilate the vacuum of the free theory
\beq
Q_{fa} | 0 \rangle = 0 
\label {Qvac=0}
\eeq
as they must since 
supersymmetry is 
unbroken and the energy of the ground state
\( | 0 \rangle \) is zero
in the free theory.

\ack
Thanks to Randolph Reeder 
for helpful conversations.
\section*{References}
\bibliography{physics}
\end{document}